\documentclass[12pt]{scrartcl}
\usepackage{graphicx}
\usepackage{natbib}

\DeclareOldFontCommand{\rm}{\normalfont\rmfamily}{\mathrm}
\DeclareOldFontCommand{\sf}{\normalfont\sffamily}{\mathsf}

\usepackage[letterpaper,left=2.5cm, right=2.5cm, top=2.5cm]{geometry}
\usepackage{url}
\usepackage{hyperref}
\usepackage{amssymb}

\setcaphanging
\setcapindent*{0pt}
\setkomafont{captionlabel}{\sffamily}

\newcommand*\aap{A\&A}

\newcommand*\aj{AJ}

\newcommand*\apj{ApJ}
\newcommand*\apjl{ApJ}

\newcommand*\mnras{MNRAS}

\newcommand*\nat{Nature}

\newcommand*\pasp{PASP}

\date{\large Submit February 2019}

\begin{document}
\sffamily
	

\raggedright
\LARGE
Astro2020 Science White Paper \linebreak

\begin{center}
	\LARGE {\textbf {Characterizing Transiting Exoplanets with JWST Guaranteed Time and ERS Observations}}\\
	\vskip 0.2 cm
\end{center}

\normalsize

\noindent \textbf{Thematic Areas:} \hspace*{59.7pt} $\blacksquare$ Planetary Systems \hspace*{2pt} $\square$ Star and Planet Formation \hspace*{20pt}\linebreak
$\square$ Formation and Evolution of Compact Objects \hspace*{29pt} $\square$ Cosmology and Fundamental Physics \linebreak
  $\square$ Stars and Stellar Evolution \hspace*{10.4pt} $\square$ Resolved Stellar Populations and their Environments \hspace*{40pt} \linebreak
  $\square$ Galaxy Evolution   \hspace*{56.5pt} $\square$ Multi-Messenger Astronomy and Astrophysics \hspace*{65pt} \linebreak
  
\textbf{Principal Author:}\\
\small
Name:	Thomas Greene
 \linebreak						
Institution:  NASA Ames Research Center
 \linebreak
Email: tom.greene@nasa.gov
 \linebreak
Phone:  650 539 5244
 \linebreak

\textbf{Co-authors:}\\
\begin{tabular}[t]{ll}
\small 
  Natalie Batalha & University of California, Santa Cruz\\
  Jacob Bean & University of Chicago\\
  Thomas Beatty & University of Arizona\\
  Jeroen Bouwman & MPIA, Heidelberg\\
  Jonathan Fortney & University of California, Santa Cruz\\
  Yasuhiro Hasegawa & JPL / Caltech\\
  Thomas Henning & MPIA, Heidelberg\\
  David Lafreni\`{e}re & University of Montreal\\
  Pierre-Olivier Lagage & CEA, Paris-Saclay University\\
  George Rieke & University of Arizona\\
  Thomas Roellig & NASA Ames Research Center\\
  Everett Schlawin & University of Arizona\\
  Kevin Stevenson & Space Telescope Science Institute
\end{tabular}
\normalsize 

\rmfamily

\section{Abstract}
We highlight how guaranteed time observations (GTOs) and early release science (ERS) will advance understanding of exoplanet atmospheres and provide a glimpse into what transiting exoplanet science will be done with $JWST$ during
its first year of operations. These observations of 27 transiting planets will deliver significant insights into the compositions, chemistry, clouds, and thermal profiles of warm-to-hot gas-dominated planets well beyond what we have learned from $HST$, $Spitzer$, and other observatories to date. These data and insights will in turn inform our understanding of planet formation, atmospheric transport and climate, and relationships between various properties. Some insight will likely be gained into rocky planet atmospheres as well. $JWST$ will be the most important mission for characterizing exoplanet atmospheres in the 2020s, and this should be considered in assessing exoplanet science for the 2020s and 2030s and future facilities.

\pagebreak
\setcounter{page}{1}

\section{Introduction} \label{sec:Intro}

There are now nearly 4000 confirmed planets listed in the NASA Exoplanet Archive. Most of these planets have been discovered via the transit technique (with $Kepler$ or otherwise), and they and their solar systems are largely different from our own. Over one-third of known exoplanets have measured masses (absolute or projected $m$ sin $i$), and the next step in understanding their compositions, formation, and climate / circulation requires spectral characterization of their atmospheres. Robust spectral characterization of a variety of planets is needed to ultimately understand their diversity, formation, and evolution and how they relate to our own Solar System and its planets. However, only a few dozen planets have been spectrally characterized reasonably well and for only a handful of atomic and molecular species, usually with high precision near-UV, visible, and near-IR spectra from $HST$, some ground-based telescopes, and $Spitzer$.\linebreak

$HST$ and other observations to date have revealed Rayleigh scattering from small particles along with atomic alkali and H$_2$O features in the transmission or emission spectra of mostly hot (T $>$ 1000 K) transiting exoplanets. There are also detections or limits on weaker molecular features (CO, CO$_2$, CH$_4$, NH3, VO, TiO, and others) as well in a number of planets \citep[e.g., see][]{2015PASP..127..941C, 2016Natur.529...59S, 2018AJ....155..156T}. In addition to providing basic composition (including any non-equilibrium chemistry) and sometimes thermal profile (vertical or longitudinal) information, these observations have provided some insight into planet formation as well \citep[e.g.,relative to the snowline as per][]{2011ApJ...743L..16O}. This significant progress has also raised numerous tantalizing questions:

\begin{itemize}
\item How does atmospheric composition vary as a function of key exoplanet properties of mass, radius, and level of insolation?
\item What are the atmospheric constituents of mini-Neptunes, super-Earths, and even terrestrial planets?
\item What can we learn about the formation of exoplanets from their C/O ratios and differences in their metallicities compared to their host stars?
\item What are the compositions of clouds and hazes in exoplanet atmospheres, under what conditions do they form, and what processes are responsible?
\item What causes chemical disequilibrium molecular abundances seen in some (mostly cooler) planets; what is the relative importance of vertical mixing, photochemistry, or other processes?
\end{itemize}

Addressing these questions will require more observations of a broader range of transiting planets and over a larger range of wavelengths.

\section{JWST atmospheric characterization capabilities}\label{sec:JWST}

$JWST$ observations will address these questions by greatly expanding the spectral characterization of transiting planets, both with regard to the wavelengths covered and to the types of transiting planets that will be accessible.
High quality data can be obtained in relatively few visits due to the large 25 m$^2$ collecting area of its primary mirror (over 5 times greater than $HST$) and its efficient operation at Earth-Sun L2. The $JWST$ instruments' time series observing modes cover a large wavelength range, $\lambda = 0.6 - 12$ $\mu$m for spectra and $\lambda = 0.6 - 26$ $\mu$m for photometric imaging, although multiple visits will be required to cover large wavelength ranges \citep{2014PASP..126.1134B}\footnote{\label{Jdox_TSO}see also \href{https://jwst-docs.stsci.edu/}{\url{https://jwst-docs.stsci.edu/}}}.
\linebreak

\begin{figure}[h]
   \centering
   \begin{minipage}[c]{0.62\textwidth}
	   \centering
	   \includegraphics[width=1.0\textwidth, angle=0]{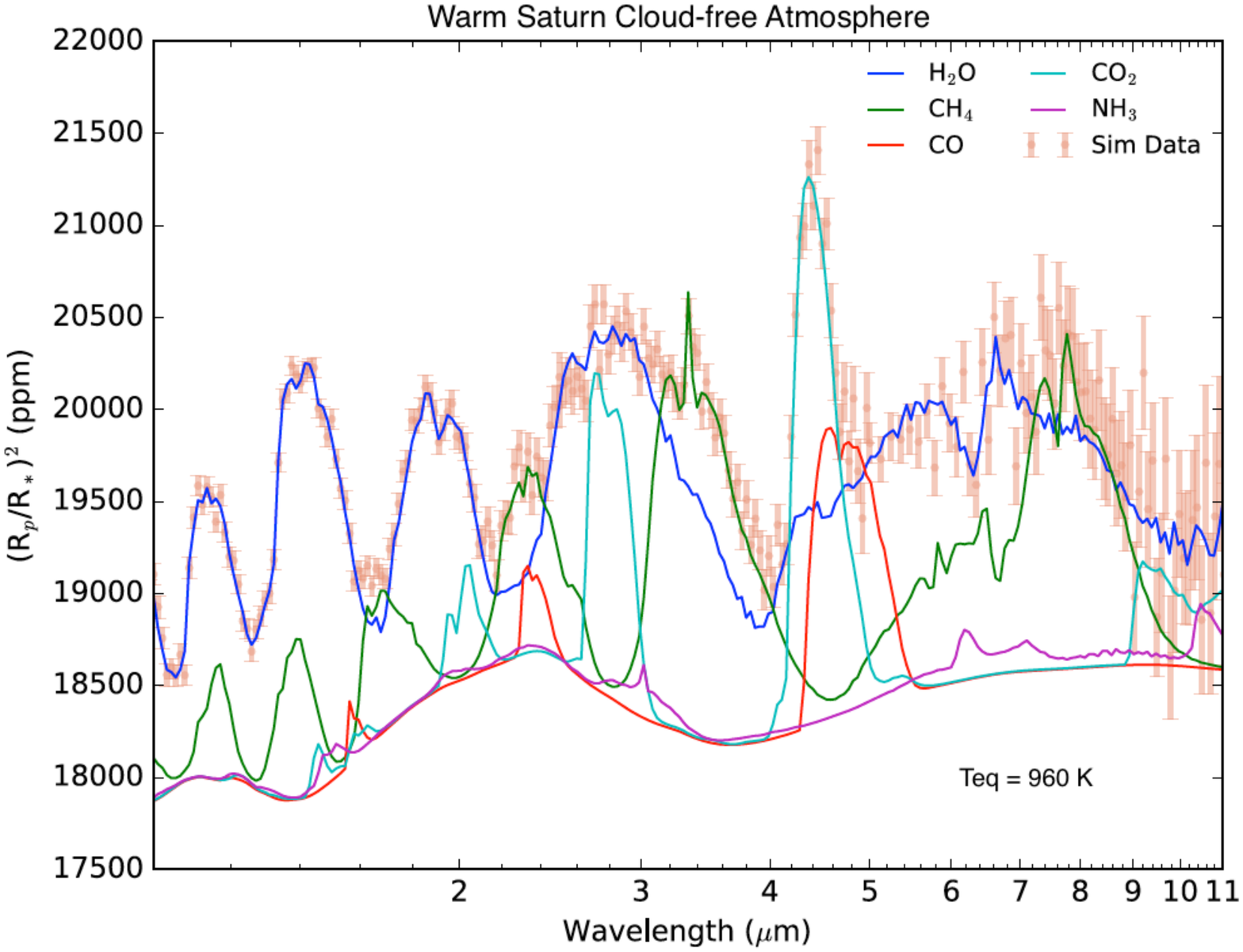}
   \end{minipage}%
   \hspace{0.03\textwidth}
   \begin{minipage}[r]{0.32\textwidth}
	   \centering
	   \caption{ \label{fig:chemistry_spec_fig}
	   \footnotesize  Equilibrium molecular feature strengths and simulated $JWST$ transmission spectrum of a warm ($T_{\rm eq} = 960$ K) sub-Saturn mass planet with a H-dominated, cloud-free  atmosphere (similar parameters to HAT-P-12 b).  Model spectra were created using the CHIMERA suite \citep{LWZ13, LKW14}, and simulated $JWST$ observations for 1 transit at each wavelength are also shown.
	   Figure provided by E. Schlawin.}
	   \end{minipage}%
\end{figure}

JWST will be able to obtain high quality transmission and emission spectra of numerous warm-to-hot planets down to $\sim$ 10 Earth masses or less \citep{2016ApJ...817...17G,2016ApJ...833..120R,2017A&A...600A..10M}, advancing exoplanet characterization into a new era.
Figure~\ref{fig:chemistry_spec_fig} shows features of H$_2$O, CH$_4$, CO, CO$_2$, and NH$_3$ that vary with temperature and will be diagnostic of the atmospheric composition and chemistry of cool to hot (400 K $\leq T_{\rm eq} \leq$ 3000 K, H-dominated planetary atmospheres. The $JWST$ spectral range is much greater than $HST$ (limited to $\lambda \leq 1.7$ $\mu$m), enabling detection of strong bands of molecules besides H$_2$O.\linebreak

\begin{figure}
	\centering
	\includegraphics[width=1.0\textwidth, angle=0]{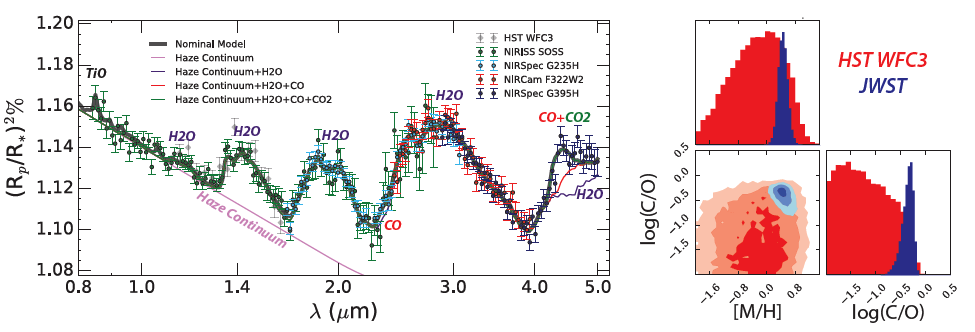}
	\caption{ \label{fig:JWST_vs_HST}
	  \footnotesize  Simulated $JWST$ transmission observations (left; one transit at each wavelength) of the hot Jupiter WASP-79 b will constrain its atmospheric C/O and [M/H] with uncertainties over an order of magnitude smaller than from $HST$ WFC3 IR data (right) as determined from CHIMERA retrievals \citep[from][]{2018PASP..130k4402B}.}
\end{figure}

Single-transit $\lambda = 1 - 2.8$ $\mu$m $JWST$ NIRISS SOSS spectra will often constrain the major molecular constituents and [M/H] of clear solar-composition atmospheres well (to factors of a few), while full $1 - 11$ $\mu$m (NIRISS + NIRCam/NIRSpec + MIRI LRS) spectra will be be needed for cloudy or high mean molecular weight atmospheres. Emission spectra will also constrain compositions and reveal the vertical pressure-temperature profiles of clear-to-cloudy atmospheres \citep{2016ApJ...817...17G}. \citet{2018AJ....156...40S} have shown that $\lambda = 1 - 2.8$ $\mu$m data are needed to supplement $\lambda > 2.4$ $\mu$m observations for measuring [M/H] with transmission spectra in some cases. Full wavelength $1 - 11$ $\mu$m data will also be essential for determining the size and composition properties of cloud components \citep[e.g.,][]{2015ApJ...815..110M}. Figure~\ref{fig:JWST_vs_HST} shows that multi-wavelength $JWST$ transmission spectra of the hot Jupiter WASP-79 b are expected to constrain its atmospheric C/O and [M/H] with 1 $\sigma$ uncertainties of $\sim$30\%, over an order of magnitude better than $HST$ WFC3 data \citep[c.f.][]{2018PASP..130k4402B}.\linebreak

Multi-wavelength $JWST$ observations of a significant sample of planet masses with a wide range of temperatures will explore the diversity of exoplanet atmospheres and will reveal relationships whose causes can be probed and tested. Sampling a range of planet masses will enable a comparison to giant Solar System planets which show increasing [M/H] with decreasing mass \citep[e.g., see][and companion white paper by J. Lunine et al.]{2014ApJ...793L..27K}. In addition to this [M/H] comparison, C/O ratios should help constrain the location of planet formation within circumstellar disks relative to snow lines \citep{2011ApJ...743L..16O}. Observing a significant number of planets with a wide range of temperatures will probe the transition from CO+CO$_2$ to CH$_4$ dominance expected at T$\lesssim$ 1000 K and sensitively detect non-equilibrium chemistry \citep[e.g.,][]{2015ApJ...813...47M, 2017A&A...600A..10M}.

\section{GTO and ERS Transit Programs}\label{sec:programs}

Approximately 3700 hours of
GTO\footnote{\label{Jdox_TSO}\href{https://jwst.stsci.edu/observing-programs/approved-gto-programs}{\url{https://jwst.stsci.edu/observing-programs/approved-gto-programs}}} and an additional $\sim$500 hours of Director's Discretionary Early Release Science\footnote{\label{ERS}\href{https://jwst.stsci.edu/observing-programs/approved-ers-programs}{\url{https://jwst.stsci.edu/observing-programs/approved-ers-programs}}} (ERS) observations have been accepted for $JWST$ Cycle 1. This is $\sim$50\% of the time available in the first year of science operations. General observer (GO) proposals will not be submitted for some time yet, so the GTO + ERS observations offer the best current glimpse of what science will be done with $JWST$ during its first year of operations.\linebreak

\begin{figure}
   \centering
   \begin{minipage}[c]{0.75\textwidth}
	   \centering
	   \includegraphics[width=1.0\textwidth, trim=50 20 50 50, angle=0]{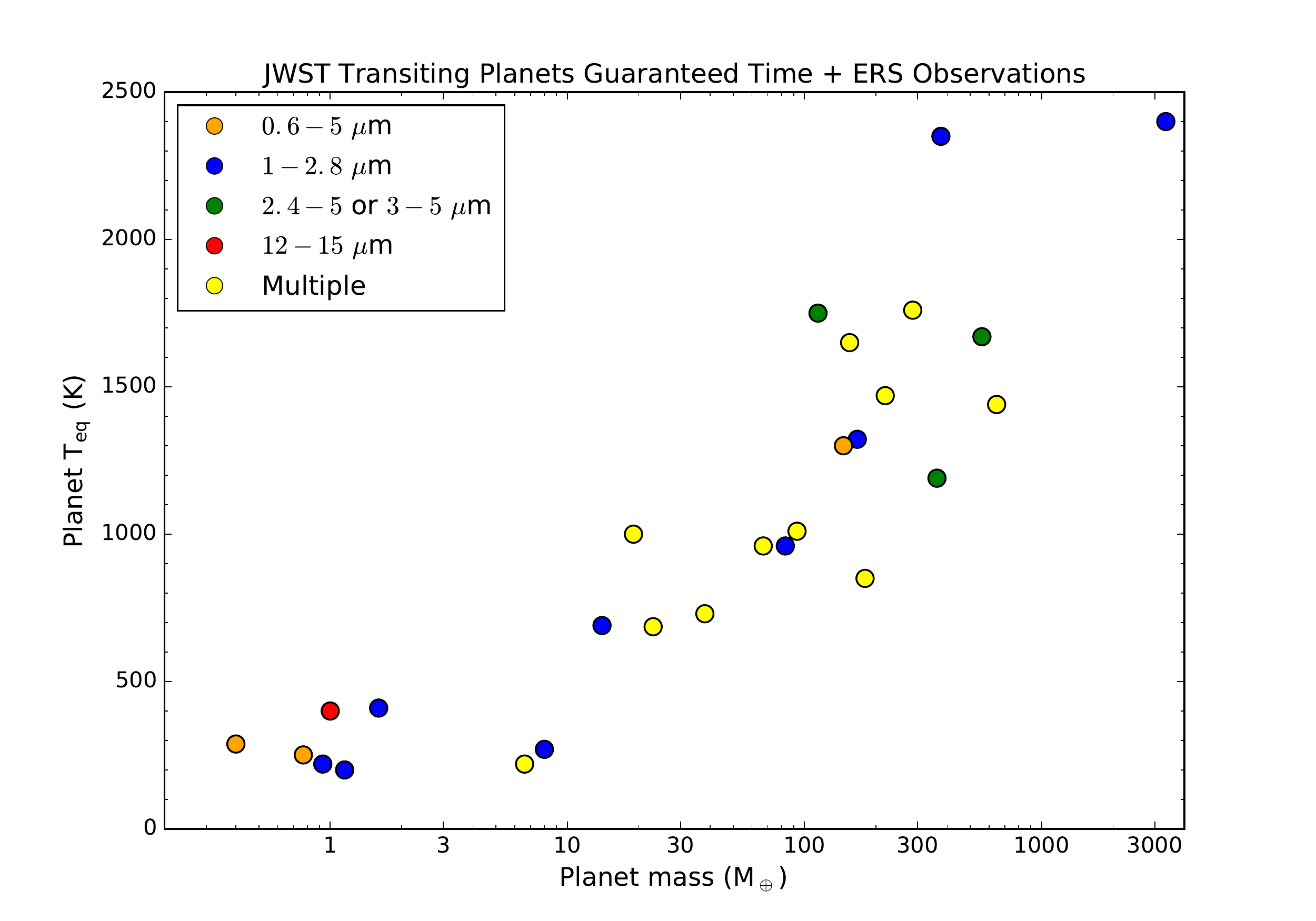}
   \end{minipage}%
   \begin{minipage}[c]{0.24\textwidth}
	   \centering
	   \caption{ \label{fig:planets}
	   \footnotesize  Transiting planets to be observed in the $JWST$ GTO and ERS programs during Cycle 1. Colors indicate wavelengths, with all but the red (MIRI) point to be made in spectroscopic modes. These include transmission, emission, and phase curve  (WASP-43 b over $3-11$ $\mu$m and WASP-121 b over $1-2.8$ $\mu$m) observations. Yellow points indicate observations with multiple instrument modes covering some combination of $\lambda = 0.6 - 5$, $1 - 2.8$, $2.4 - 5$, $3 - 5$, and $5 - 11$ $\mu$m.
	   }
	   \end{minipage}%
\end{figure}

The transiting planet observations in the Cycle 1 GTO and ERS \citep{2018PASP..130k4402B} programs will enable a large step forward in the characterization of exoplanet atmospheres. Table~\ref{table:programs} lists these programs which sum up to 816 hours, 19\% of the scheduled GTO+ERS observing time. Figure \ref{fig:planets} shows the diversity of the masses and equilibrium temperatures of the 27 planets in these programs as well as the wavelengths of their observations. A companion white paper by C. Beichman et al. describes additional GTO+ERS resolved imaging and spectroscopy exoplanet programs.\linebreak
	
\begin{table}
    \centering
  \small
\caption{\bfseries Approved GTO and ERS Transiting Planet Programs}
\begin{tabular}{cllr}
{\bfseries ID}&{\bfseries Title and Science Instrument}&{\bfseries Team Lead}&{\bfseries Hours}\\ \hline
1177 & MIRI observations of transiting exoplanets & T. Greene & 75 \\
1185 & Transit Spectroscopy of Mature Planets (NIRCam) & T. Greene & 140 \\
1201 & NIRISS Exploration of the Atmospheric Diversity of Transiting Exoplanets  & D. Lafreni{\`e}re & 201 \\
1224 & Transiting Exoplanet Characterization with JWST/NIRSPEC & S. Birkmann & 50 \\
1274 & Extrasolar Planet Science with $JWST$ (NIRCam) & J. Lunine & 74 \\
1279 & Thermal emission from Trappist1-b (MIRI) & P.-O. Lagage  & 25 \\
1280 & MIRI Transiting Observation of WASP-107b & P.-O. Lagage  & 11 \\
1281 & MIRI and NIRSPEC Transit Observations of HAT-P-12 b & P.-O. Lagage  & 32 \\
1312 & Transit and Eclipse Spectroscopy of a Warm Neptune (NIRISS+NS+MIRI) & N. Lewis & 36 \\
1331 & Transit Spectroscopy of TRAPPIST-1e (NIRSpec) & N. Lewis & 22 \\
1353 & Transit and Eclipse Spectroscopy of a Hot Jupiter (NIRISS+NS+MIRI) &  N. Lewis & 72 \\
1366 & The Transiting Exoplanet Community ERS Program (all SIs) & N. Batalha & 78 \\
     & \bfseries{TOTAL}  &  & 816\\
\hline \hline
\end{tabular}
\label{table:programs}
\end{table}

Some changes may still be made to the GTO and ERS programs. The current list of all current GTO transiting planet observations is available here: \href{https://goo.gl/W9Q7wY}{\url{https://goo.gl/W9Q7wY}}.


\section{Conclusions}

$JWST$ data will usher in a new age of exoplanet atmosphere characterization, enabling precision measurements of molecular abundances, metallicities, temperature profiles, and chemistry. These results will inform our understanding of planet formation, star-planet interactions, and circulation and chemistry within planetary atmospheres.\linebreak

The GTO and ERS programs approved for Cycle 1 include multi-wavelength transmission, emission, and phase curve observations of 27 transiting planets with masses from under 1 M$_\oplus$ to 10 times the mass of Jupiter and temperatures from 200 -- 2400 K. The not-yet-selected Cycle 1 GO programs will consume a similar amount of total time as all GTO+ERS, so $JWST$ may well characterize the atmospheres of over 50 transiting planets in its first year of science operations. This is significantly more than the ones studied in detail so far with $HST$ and $Spitzer$, which have mostly been limited to more massive worlds. The higher precision and larger wavelength coverage of JWST will enable the comprehensive characterization that is needed to advance our understanding of planet formation and atmospheric processes in the 2020s.\linebreak

This combination of scientific capability, sensitivity to diverse planets, and relatively efficient survey capability will make $JWST$ the most important observatory for exoplanet characterization in the 2020s. Despite this incredible power, $JWST$ is expected to make only modest progress in characterizing atmospheres of Earth-like worlds in habitable zones \citep{2016MNRAS.458.2657B,2016ApJ...819L..13S}. High contrast direct imaging will likely be needed to find and characterize a significant number of habitable-zone Earth-like worlds. Astro2020 should consider this when assessing future science opportunities and recommending the facilities and missions needed to realize them.\linebreak

\vfill\eject

\end{document}